\documentstyle[emulateapj,epsfig]{article}

\begin{document}
\submitted{ApJ in press}

\title{Formation of the First Supermassive Black Holes}

\author{Volker Bromm\altaffilmark{1,4} 
and Abraham Loeb\altaffilmark{1,2,3,4}}


\altaffiltext{1} {Astronomy Department, Harvard University, 60 Garden
St., Cambridge, MA 02138 }

\altaffiltext{2}{Institute for Advanced Study, Princeton, NJ 08540}

\altaffiltext{3}{Guggenheim Fellow}

\altaffiltext{4}{Email: vbromm@cfa.harvard.edu; loeb@ias.edu}

\begin{abstract}
We consider the physical conditions under which supermassive black
holes could have formed inside the first galaxies. Our SPH simulations
indicate that metal-free galaxies with a virial temperature $\sim
10^4$ K and with suppressed H$_2$ formation (due to an intergalactic
UV background) tend to form a binary black hole system which contains
a substantial fraction ($\ga 10\%$) of the total baryonic mass of the
host galaxy. Fragmentation into stars is suppressed without
substantial H$_2$ cooling.  Our simulations follow the condensation of
$\sim 5\times 10^6 M_\odot$ around the two centers of the binary down
to a scale of $\la 0.1$~pc.  Low-spin galaxies form a single black
hole instead.  These early black holes lead to quasar activity before
the epoch of reionization.  Primordial black hole binaries lead to
gravitational radiation emission at redshifts $z\ga 10$ that would be
detectable by LISA.

\end{abstract}

\section{INTRODUCTION}

Supermassive black holes (SMBHs) are believed to provide the power
source of quasars via the accretion of surrounding gas (Salpeter 1964;
Zeldovich 1964; Lynden-Bell 1969; Rees 1984).  They were able to form
already at early cosmic times, as implied by the recent discovery of
quasars at redshifs $z\ga 6$ (Becker et al. 2001; Djorgovski et
al. 2001; Fan et al. 2002).  The existence of SMBHs with inferred
masses of $\ga 10^{9}M_{\odot}$, less than a billion years after the
big bang, provides important constraints on any formation scenario
(Haiman \& Loeb 2001).

The gas physics involved in the formation of SMBHs is still
not well understood (Rees 1984, 2002; Loeb \& Rasio 1994; Barkana \& Loeb
2001).  In this work, we attempt to simulate this process under the
simplest and best prescribed set of initial conditions; those dictated
by the early universe.  The earliest galaxies are simple in that they
are the first condensations of gas to grow out of the seed
inhomogeneities in the early universe.  The composition of the
primordial gas is determined by big bang nucleosynthesis (Burles,
Nollett, \& Turner 2001, and references therein) and any primordial
magnetic fields are not expected to be dynamically significant.

Previous numerical simulations of early luminous structures (Bromm,
Coppi, \& Larson 1999, 2002; Abel, Bryan, \& Norman 2000, 2002;
Nakamura \& Umemura 2001) have focused on the formation of stars
inside the very first gaseous objects with a mass, $\sim 10^5M_\odot$,
just above the cosmological Jeans mass.  Due to their low virial
temperature (hundreds of K), fragmentation into stars is possible
inside these objects through the formation of molecular hydrogen,
H$_2$, which cools efficiently via rotational-vibrational transitions
even at these low temperatures. However, a relatively modest UV flux
is sufficient to photo-dissociate the fragile H$_2$ molecules and
suppress their role in cooling the gas (Haiman, Rees, \& Loeb 1997;
Ciardi et al. 2000; Haiman, Abel, \& Rees 2000).  A destructive flux
of UV photons could be produced by a small, early population of stars.

Here we consider dwarf galaxies with virial temperatures $\ga 10^4$~K
inside of which atomic cooling is effective (Oh \& Haiman 2002).
These galaxies comprise a substantial population at $z\sim 10$
(Barkana \& Loeb 2001) that is expected to dominate the reionization
of hydrogen at $z\ga 6$ (Wyithe \& Loeb 2003a).  We use smoothed
particle hydrodynamics (SPH) simulations to describe the cooling and
dynamics of the gas for different choices of its total angular
momentum (or spin parameter), following the suggestion by Eisenstein
\& Loeb (1995) that low-spin systems would be more susceptible to the
formation of a SMBH.  If H$_2$ cooling is suppressed inside these
galaxies, then their gas will not cool below $10^4$K. When the
temperature to which the gas cools is only somewhat lower than the
virial temperature of the host galaxy (at which the pressure force
balances gravity for the gas), we expect that fragmentation into small
clumps will be avoided and the gas will tend to condense isothermally
(at its temperature floor of $\sim 10^4$K) into large clumps. Such
large clumps may then collapse to form a SMBH, possibly through an
intermediate stage of a supermassive star.  The viability of this
scenario relies on the suppression of molecular H$_2$ cooling, which
when present is capable of cooling the gas to a temperature as low as
200 K.  We therefore start our analysis in \S~2 by considering the
cosmic UV background expected from the first stars. Later on, in \S~5
we carry out numerical simulations to determine the level of UV
radiation necessary for the photodissociation of H$_2$ molecules
during the early collapse phase of the above dwarf galaxies.

The SPH code we use is the same as was used to simulate the formation
of the first massive stars (Bromm et al. 1999, 2002), except that we
apply it to more massive galaxies where atomic cooling is effective.
Our goal is to find whether there is a direct route to the formation
of SMBHs under these circumstances.  Although our
focus is on gas dynamics, we note that the first massive black holes (BHs)
may have also formed out of the gravitational dynamics of clusters of
stellar-mass BHs (Larson 2000, 2002; Madau \& Rees 2001; Schneider et al. 2002;
Islam, Taylor, \& Silk 2003; Volonteri, Haardt, \& Madau 2003).

Throughout this paper, we assume a standard $\Lambda$CDM cosmology,
with a total density parameter in matter of $\Omega_{m}=1-
\Omega_{\Lambda}=0.3$, and in baryons of $\Omega_{\rm B}=0.045$.  The
Hubble constant is $h=H_{0}/(100$ km s$^{-1}$ Mpc$^{-1}$)=0.7, and the
present-day power-spectrum amplitude is $\sigma_{8}=0.9$ in spheres of
radius $8h^{-1}$Mpc.

\section{THE COSMOLOGICAL CONTEXT}

We first outline the basic physical reason for why small protogalactic
systems of virial temperature $\sim 10^4$K and mass $\sim
10^{8}M_{\odot}$, collapsing before the epoch of reionization at
redshifts $z\ga 10$, provide intriguing sites for the formation of the
first SMBHs.

The direct collapse of a primordial gas cloud into a central compact
object is made difficult by fragmentation and consequent star
formation (Loeb \& Rasio 1994). The ability of the gas to fragment
depends on the presence of an efficient cooling mechanism.  By a
redshift of $z\sim 10$, the intergalactic medium (IGM) is expected to
be enriched with heavy elements from the first generation of stars to
a level of $Z\ga 10^{-3.5}Z_{\odot}$ (e.g., Gnedin \& Ostriker 1997;
Mackey, Bromm, \& Hernquist 2003). Recent numerical investigations
have shown that such a metallicity is sufficient to enable the gas to
efficiently cool, fragment, and subsequently form stars (Omukai 2000;
Bromm et al.  2001a; Bromm \& Clarke 2002). 
It is plausible that in some regions of the IGM, the
gas that collapses into a dwarf galaxy has not been enriched with
metals in excess of $Z_{\rm crit}\sim 10^{-3.5}Z_{\odot}$ (e.g.,
Thacker, Scannapieco, \& Davis 2002; Furlanetto \& Loeb 2003).  In
this paper, we investigate the collapse of a virtually metal-free
system of this type.

It is well known that molecular hydrogen can effectively cool
metal-free gas, and thus enable the formation of the first stars at
$z\simeq 20-30$ (e.g., Haiman, Thoul, \& Loeb 1996; Tegmark et
al. 1997). Even in the absence of metals, star formation is therefore
expected to occur, and the subsequent negative feedback due to
supernova explosions would prevent the assembly of large quantities of
gas in the center of the shallow dark matter (DM) potential well that
characterizes the first dwarf galaxies (e.g., Dekel \& Silk
1986). Molecular hydrogen, however, is fragile and can readily be
destroyed by photons in the Lyman-Werner (LW) bands, within the energy
range $11.2-13.6$~eV, via the two-step Solomon process (Stecher \&
Williams 1967)
\begin{displaymath}
{\rm H}_{2} + \gamma \rightarrow {\rm H}_{2}^{\ast} \rightarrow 2 {\rm H} {\rm .}
\end{displaymath}
The intermediate stage involves an excited electronic 
state, ${\rm H}_{2}^{\ast}$, from which a fraction of the subsequent
decays end in the vibrational continuum of the ground state, resulting
in the dissociation of the molecule.

The question then arises whether H$_{2}$ cooling can indeed be
suppressed in the pre-reionization dwarf galaxies considered in this
paper. These systems are rather massive compared to the $\sim 10^{5} -
10^{6}M_{\odot}$ halos that host the formation of the first stars at
$z\ga 20$. The gas might then be able to self-shield against the
photo-dissociating LW background (e.g., Glover \& Brand 2001;
Machacek, Bryan, \& Abel 2001).  Close to the epoch of reionization,
however, a significant flux in the LW bands ($h\nu < 13.6$~eV) is
expected. To estimate the LW flux, we first consider the flux of
ionizing radiation just above the Lyman limit,
\begin{equation}
J^{+}_{\nu}\simeq \frac{h c}{4\pi}\frac{N_\gamma X \rho_{\rm
B}(z=10)}{m_{\rm H}} \mbox{\ .}
\end{equation}
Here, $\rho_{\rm B}(z=10)$ is the baryonic density at $z\simeq 10$,
$X=0.76$ the mass fraction in hydrogen, and $m_{\rm H}$ the mass of
a hydrogen atom.  We assume that $N_\gamma\sim 10$ ionizing photons per baryon
are required to reionize the universe (Wyithe \& Loeb
2003a). Evaluating this expression for the cosmological parameters
given above, we find: $J^{+}_{21}\sim 40$.  Here, and in the remainder
of the paper, UV fluxes are normalized as $J_{\nu}=J_{21}\times
10^{-21}$~erg cm$^{-2}$~s$^{-1}$~Hz$^{-1}$~sr$^{-1}$.

The flux in the LW bands just below the Lyman limit, $J_{\nu}^{-}$
could, however, be much larger than $J_{\nu}^{+}$. Assuming that only a
fraction, $f_{\rm esc}$, of the ionizing photons can escape from the
star forming halos, we have: $J^{-}_{21}\simeq J^{+}_{21}/f_{\rm
esc}\sim 4\times 10^3 (f_{\rm esc}/0.01)$, where $f_{\rm esc}$ is
expected to be low at high redshifts (Wood \& Loeb 2000, and
references therein)\footnote{A strong background of photo-dissociating
photons could also result from star formation in the dwarf galaxy
itself (e.g., Omukai \& Nishi 1999; Oh \& Haiman 2002). However, this
would also result in the production of metals, probably leading to the
rapid enrichment of the galaxy beyond $Z_{\rm crit}$.}.

The strong UV background flux estimated above makes it possible for
H$_{2}$ formation to be effectively suppressed close to the redshift
of reionization.  In the absence of molecular hydrogen, however,
cooling can still proceed via atomic transitions in halos of mass (see
Barkana \& Loeb 2001)
\begin{equation}
M\ga
 10^{8} M_{\odot} \left(\frac{1+z}{10}\right)^{-3/2}
\mbox{\, .}
\end{equation}
The virial temperature, $T_{\rm vir}\sim 10^{4}$K, in these more
massive halos allows for the very efficient cooling of the gas via
lines of atomic hydrogen.  Notice that in this case where the gas
temperature is close to the virial temperature, the gas cloud as a
whole can undergo collapse but it will not be able to fragment until
high enough densities are reached so that the Jeans mass has declined
sufficiently.  A halo of total mass $\sim 10^{8}M_{\odot}$ and a
collapse redshift $z\sim 10$ corresponds to a 2$\sigma$~peak in the
random field of primordial density fluctuations.

In \S~4, we describe numerical simulations of such a 2 $\sigma$~peak.
We consider the evolution with and without the presence of H$_{2}$.
To ascertain the importance of H$_{2}$, we consider the limiting cases
of a halo made of purely atomic hydrogen, and of a halo where H$_{2}$
is allowed to form without any negative feedback.  Subsequently, in
\S~5 we discuss simulations where an external LW background is
included. We in particular assess the critical LW flux required to
prevent molecules from forming.  This required background flux may
then be compared to the available flux at $z\sim 10$, as estimated in
this section.

\setcounter{figure}{0}

\section{Numerical Methodology}
\subsection{Large-scale Simulations}


\begin{deluxetable}{lcccccc} 
\footnotesize
\tablewidth{11.cm}
\tablecaption{Parameters for the different Runs \label{tab1}}
\tablecolumns{7}
\tablehead{
\colhead{Run} &
\colhead{$M$} &
\colhead{$z_{\rm vir}$} &  
\colhead{$\lambda$} & 
\colhead{H$_{2}$}  &
\colhead{LW}  &
\colhead{$n_{\rm th}$}  \\ 
\colhead{} &
\colhead{($M_{\odot}$)} &
\colhead{} &  
\colhead{} & 
\colhead{}  &
\colhead{}  &
\colhead{(cm$^{-3}$)}  
 } 
\startdata
A ...... & $10^{8}$ & 10 & 0 & No & No & $10^{7}$ \nl
B ...... & $10^{8}$ & 10 & 0.05 & No & No & $10^{7}$ \nl
C ...... & $10^{8}$ & 10 & 0 & Yes & No & $10^{4}$ \nl
D ...... & $10^{8}$ & 10 & 0 & Yes & Yes & $10^{7}$ \nl
\enddata
\tablecomments{$M$ is the total mass of the halo, $z_{\rm vir}$ the 
collapse redshift, $\lambda$ the spin parameter, 
$n_{\rm th}$ the threshold density for the creation of a sink particle, 
H$_{2}$ refers to the
absence or presence of molecular hydrogen cooling, and LW denotes whether
a photodissociating Lyman-Werner background is included or not.}
\end{deluxetable}

The evolution of the dark matter and gas components is calculated with
a version of TREESPH (Hernquist \& Katz 1989), combining the
SPH method with a hierarchical (tree) gravity
solver (see Bromm et al. 2002 for further details).  Here, we briefly
describe the additions to the code which are necessary for the
investigation of zero-metallicity gas.  These include a method to treat
the radiative cooling of the gas together with a multi-species chemical
reaction network, and a technique to create sink
particles.  The thermal evolution of the gas is governed by the
equation:

\begin{equation}
\frac{{\rm D}u}{{\rm D}t}=\frac{P}{\rho^{2}}\frac{\rm{D}\rho}{\rm{D}t}
+ \frac{\Gamma-\Lambda}{\rho}
\end{equation}
 where
 ${\rm D}/{\rm D}t$ is the Lagrangian time derivative,
 $P$ and $\rho$ are the gas pressure and density, $u$ is the specific
internal energy (in erg g$^{-1}$),
and $\Gamma$ and $\Lambda$ are the contributions from radiative heating and
cooling, respectively (in erg cm$^{-3}$ s$^{-1}$).
Since we are concerned with a system that formed prior to the epoch of
reionization, we assume that there is no heating due to a photoionizing
background ($\Gamma =0$). Radiative cooling is due to lines of atomic 
hydrogen at gas temperatures $T\ga 10^{4}$~K, and lines of molecular
hydrogen at lower temperatures.
We have implemented the
H$_{2}$ cooling function given by Galli \& Palla (1998).
Finally, the first term on the right-hand side describes adiabatic
cooling due to expansion or heating due to compression.

Since radiative cooling to temperatures below that of the cosmic microwave
background (CMB), $T_{\rm CMB}=2.7\mbox{\,K}(1+z)$, is thermodynamically
not possible, we write the cooling term as
\begin{equation}
\Lambda=\Lambda(T)-\Lambda(T_{\rm CMB})\mbox{\ \ \ .}
\end{equation}
For $T < T_{\rm CMB}$, radiative cooling consequently turns into heating.
This approximate treatment ensures that $T\geq T_{\rm CMB}$, unless cooling proceeds via
adiabatic expansion.

The chemical reaction network comprises the 9 species H, H$^{+}$, H$^{-}$,
H$_{2}$, H$_{2}^{+}$, $e^{-}$, He, He$^{+}$, and He$^{++}$, including the
reactions given in Haiman et al. (1996). Our implicit, backwards-differencing
method to solve the coupled set of rate equations is fast and accurate
(see Bromm et al. 2002 for test calculations).

We have devised an algorithm to merge SPH particles in high-density regions
in order to overcome the otherwise prohibitive time-step limitation, as
enforced by the Courant stability criterion. To follow the simulation
for a few dynamical times, we allow SPH particles to merge into more
massive ones, provided they exceed a predetermined density threshold.
More details of the merging algorithm are given in Bromm et al. (2002).

\subsection{Small-scale Simulations}

The sink-particle technique allows one to study the collective
dynamics of multiple centers of condensation, such as merging between
clumps and ongoing accretion over many dynamical timescales (e.g.,
Bate, Bonnell, \& Price 1995; Bate, Bonnell, \& Bromm 2003).
It is important, however, to also be able to follow the collapse to
increasingly high density.  To accomplish this, we have carried out a
complementary simulation in which we do not create sink
particles. Instead, we focus on the highest-density region in one of
the large-scale simulations, resample the density field with an
increased number of SPH particles, and adopt a rapidly decreasing
timestep according to $\Delta t_{\rm sys}\propto 1/\sqrt{G\rho_{\rm max}}$. 
Here, $\rho_{\rm max}$ is the maximum gas density at a given instant, and
$\Delta t_{\rm sys}$, the system timestep, is the maximum allowed time 
by which a particle is advanced within the multiple-timestep scheme
employed in the simulations (see Hernquist \& Katz 1989).
In doing so, the overall system does hardly evolve at all, while the
runaway collapse of one small region proceeds on a sufficiently short
timescale.  The mass resolution of a simulation is approximately
\begin{equation}
M_{\rm res}=\left(\frac{N_{\rm neigh}}{N_{\rm SPH}}\right)M_{\rm B}
\mbox{\ \ \ ,}
\end{equation}
where $N_{\rm neigh}\simeq 32$ is the number of particles within a given
SPH smoothing kernel, $N_{\rm SPH}$ the total number of SPH particles,
and $M_{\rm B}$ the total baryonic mass. To avoid numerical
fragmentation, the Jeans scale has to be resolved (Bate \& Burkert
1997): $M_{\rm res}<M_J$.  After the onset of gravitational instability in
the large-scale simulation, the rapid increase in density leads to a
violation of this criterion. At this point, a sink particle is
created in the simulation.

In our refined simulation, we follow a different computational
strategy (see also Bromm 2000).  Our starting configuration is the
large-scale simulation, stopped at a moment briefly before the central
region undergoes runaway collapse.  We now apply the following
resampling procedure to the fluid within this region. Every SPH
particle in the original, unrefined simulation acts as a parent
particle denoted by $p$, and spawns $N_{\rm ref}$ child particles
denoted by $k$, where $N_{\rm ref}=50$ in this paper. The child
particles are distributed according to the SPH smoothing kernel
$W(\vec{r}_{k}-\vec{r}_{p}; h_{p})$, by employing a standard
Monte-Carlo comparison-rejection method (see, e.g., Press et
al. 1992). Here, $h_{p}$ is the smoothing length of the parent
particle. The velocity, temperature, and fractional abundances of the
9 species included in the reaction network are directly inherited from
the parent particle, $\vec{v}_{k}=\vec{v}_{p}$, $T_{k}=T_{p}$, and
$y_{k}=x_{p}$ for all $k$, respectively. Finally, each child particle
is assigned a mass of $m_{k}=m_{p}/N_{\rm ref}$. This procedure conserves
linear and angular momentum well. Energy is also
conserved well, although there arises a small artificial
contribution to the gravitational potential energy due to the
discreteness of the resampling.  The resampling described in \S~4.2
results in $N_{\rm SPH}=$ 65,500 within the central high-resolution
region, and the mass resolution is now $M_{\rm res}\simeq 500 M_{\odot}$,
as compared to $M_{\rm res}\simeq 20,000 M_{\odot}$ in the original
simulation.

As we follow our simulation to densities in excess of $n_{\rm H}\sim
10^{7}$~cm$^{-3}$, opacity effects begin to modify the thermal
behavior of the gas. For simplicity, we continue to use the optically
thin rate for cooling due to the collisional excitation of atomic
hydrogen lines (Cen 1992).  While this is clearly an
oversimplification, we have verified that the resulting thermal
evolution of the gas is very similar to what is found in a more
sophisticated treatment of the cooling processes at high densities
(Omukai 2001).

Our technique of refining a coarser, parent simulation, and following
the further collapse with increased resolution, is conceptually
similar to the Adaptive Mesh Refinement (AMR) method which was
originally developed by Berger \& Oliger (1984).  The AMR method has
already been successfully applied to astrophysical problems (e.g.,
Truelove et al. 1998; Norman \& Bryan 1999), but our approach is one
of the first attempts of implementing such a scheme within SPH (see
also Kitsionas \& Whitworth 2002).

\subsection{External UV background}

To investigate the effect of an external UV background on the
evolution of the gas, we have included the relevant photo-reactions
into our chemical network. The photo-rates (in units of s$^{-1}$) 
are given by
\begin{equation}
k_{\rm ph}=4\pi\int_{\nu_{\rm th}}^{\infty}
\frac{\sigma_{\nu}J_{\nu} \mbox{d}\nu}{h\nu}\mbox{\ ,}
\end{equation}
where $h\nu_{\rm th}$ is the threshold energy for a specific reaction.
The cross-sections are given in Haiman et al. (1996). Following 
Omukai (2001), we consider both a power-law spectrum, $J_{\nu}\propto
\nu^{-1}$, appropriate for quasar sources, and a thermal Planck
spectrum, $J_{\nu}\propto B_{\nu}(T_{\ast})$, appropriate
for stellar sources.
In the latter case, we evaluate the photo-rates for two different
values of the radiation temperature: $T_{\ast}\simeq 10^{4}$~K and
$10^{5}$~K, corresponding to `normal' Population~II stars and
to very massive Population~III stars, respectively (Bromm, Kudritzki,
\& Loeb 2001b). For each spectrum, we normalize the flux at the
Lyman limit with a normalization constant $J_{21}$, and assume zero
flux at $h\nu > 13.6$~eV due to intergalactic HI absorption.

To take into account the effect of self-shielding, we write the
H$_{2}$ photodissociation rate as $k_{\rm diss}\propto J_{21}f_{\rm
shield}$, with a proportionality constant that depends on the chosen
spectrum. For the shielding factor, we use the approximate expression
$f_{\rm shield}\simeq \min[1,(N_{\rm H_{2}}/10^{14}{\rm
cm}^{-2})^{-0.75}]$ (Draine \& Bertoldi 1996). This formula is
accurate only for a static medium, and it will grossly overestimate
the H$_{2}$ line opacity in the presence of large-scale velocity
gradients. In our simulations, we encounter moderately strong bulk
flows, with velocities of order the sound speed.  We therefore
conservatively overestimate the effect of self-shielding.  The H$_{2}$
column density is estimated from local quantities only as follows:
$N_{\rm H_{2}}\simeq 0.1 f_{\rm H_{2}} n_{\rm H} L_{\rm char}$. We
define a local characteristic length, $L_{\rm char}$, such that the
total baryonic mass would be contained in a sphere of uniform density
$n_{\rm H}$, and radius $L_{\rm char}$. Thus, we have
\begin{equation}
L_{\rm char}=\left(\frac{3 X M_{\rm B}}{4\pi m_{\rm H}n_{\rm H}}\right)^{1/3}
\mbox{\ .}
\end{equation}
By comparing with the column densities that are actually realized
in the simulations, we have verified the validity of this prescription.

We neglect any contributions to the heating of the gas due to
photo-reactions, since these are always negligible in the system
studied here. We also neglect, for simplicity, the sawtooth-modulation
of the background spectrum due to the IGM line-opacity in the Lyman
series (Haiman et al. 1997).

\subsection{Initial Conditions}

Within the hierarchical $\Lambda$CDM model, the first luminous objects
are expected to form out of high $\sigma$ peaks in the random field of
primordial density fluctuations.  The early (linear) evolution of such
a peak, assumed to be an isolated and roughly spherical overdensity, can be
described by the top-hat model (e.g., Padmanabhan 1993). We use the
top-hat approximation to specify the initial DM configuration.  In
this paper, we investigate the fate of a 2 $\sigma$ peak of total mass
$10^{8}M_{\odot}$, corresponding to $1.5\times 10^{7}M_{\odot}$ in
baryons.  On this mass scale, one can estimate the redshift of
collapse (or virialization) to be $z_{\rm vir}\simeq 10$.

Our simulation is initialized at $z_{i}=100$, by performing the
following steps.  The collisionless DM particles are placed on a
cubical Cartesian grid, and are then perturbed according to a given
power spectrum $P(k)=A k^{n}$, by applying the Zeldovich (1970)
approximation which also allows to self-consistently assign initial
velocities. The power-law index is set to $n=-2.5$ which approximately
describes the spectral behavior on the scale $\sim 10^{8}M_{\odot}$.
To fix the amplitude $A$, we specify the initial variance of the
fluctuation amplitude
\begin{equation}
\sigma_{i}^{2}=A\sum k^{n}
\mbox{\ \ \ .}
\end{equation}
The summation is over all contributing modes, where the minimum
wavenumber is given by the overall size of the Cartesian box, and the
maximum wavenumber by the Nyquist frequency. Choosing
$\sigma_{i}^{2}\simeq 0.01$, the rms amplitude of fluctuations at the
moment of collapse is approximately
\begin{equation}
\sigma(z=10)\simeq\left(\frac{1+z_{i}}{1+z}\right)\sigma_{i}\simeq 1
\mbox{\ \ \ .}
\end{equation}
This choice ensures that the substructure develops on a similar
timescale as the overall collapse of the background medium.  Next,
particles within a (proper) radius of $R_{i}=$ 776 pc are selected for
the simulation. The resulting number of DM particles is $N_{\rm
DM}=17074$.  Finally, the particles are set into rigid rotation and
are endowed with a uniform Hubble expansion (see also Katz 1991).
Angular momentum is added by assuming a spin parameter
$\lambda=L|E|^{1/2}/(G M^{5/2})=0$ and 0.05, where $L$, $E$, and $M$
are the total angular momentum, energy, and mass, respectively. The
spin parameter is a measure of the degree of rotational support, such
that the ratio of centrifugal to gravitational acceleration is given
by $\sim \lambda^{2}$ at virialization. The second value corresponds
to the average spin parameter found in cosmological simulations (e.g.,
Barnes \& Efstathiou 1987;  Jang-Condell \& Hernquist 2001).

The collisional SPH particles ($N_{\rm SPH}=32768$) are randomly
placed to approximate a uniform initial density.  The SPH particles
follow the same Hubble expansion and rigid rotation as the DM
particles.  For the initial gas temperature at $z_{i}=100$ we adopt the
value of $200$ K (Tegmark et al. 1997). The fractional free-electron
abundance is initialized as $x_{e}=4.6\times 10^{-4}$, and the hydrogen
molecule abundance, in the runs where H$_{2}$ formation is allowed, as
$f_{\rm H_{2}}=2\times 10^{-6}$ (Anninos \& Norman 1996).

In Table~1, we summarize the parameters of the large-scale
simulations. We will discuss the initial state of the refined,
small-scale simulation in \S~4.2.  We empasize that the initial
particle setup, and in particular the realization of the DM
fluctuations, is identical for all runs which only differ in the
choice of spin, in the presence or absence of a soft UV background,
and in whether H$_{2}$ cooling is allowed or not. The runs with no
H$_{2}$ cooling correspond to the limiting case in which all the
H$_{2}$ has been radiatively destroyed by a soft UV background.  In
\S~2, we have discussed the possible emergence of such an efficient
feedback following the build-up of the cosmic UV background close to
the epoch of reionization.

\section{The Simulations}

In this section, we only consider simulations where no external 
UV background is included, and defer the discussion of its effect
to \S~5.
We carry out these simulations in two separate steps. First, we simulate
the large-scale evolution on the scale of the host halo, $\sim 1$~kpc,
ending with the formation of high-density clumps on a scale of $\sim
1$~pc.  Once the mass resolution of the simulation becomes larger than
the local Jeans mass, $M_{J}\la M_{\rm res}$, a sink particle is
created. Consequently, the internal dynamics of a clump cannot be
studied any further in these coarse-grain simulations. Second, we
follow the evolution of a clump to higher densities and smaller
spatial scales, by refining the resolution in the vicinity of the
clump to again resolve the Jeans mass.  We begin with a discussion of
the large-scale evolution.

\subsection{Large-scale evolution}

We first describe the run with no spin and with suppressed H$_{2}$
formation (Run A) in more detail, and subsequently explore what
happens under variations of these assumptions. In Figure 1, we show the
initial configuration at $z_{i}=100$, which is identical for all runs.
The overdense region initially expands with the Hubble flow
at a reduced rate, subsequently turns around at $z_{\rm ta}\sim 16$,
and eventually collapses.

\begin{center} 
\epsfig{file=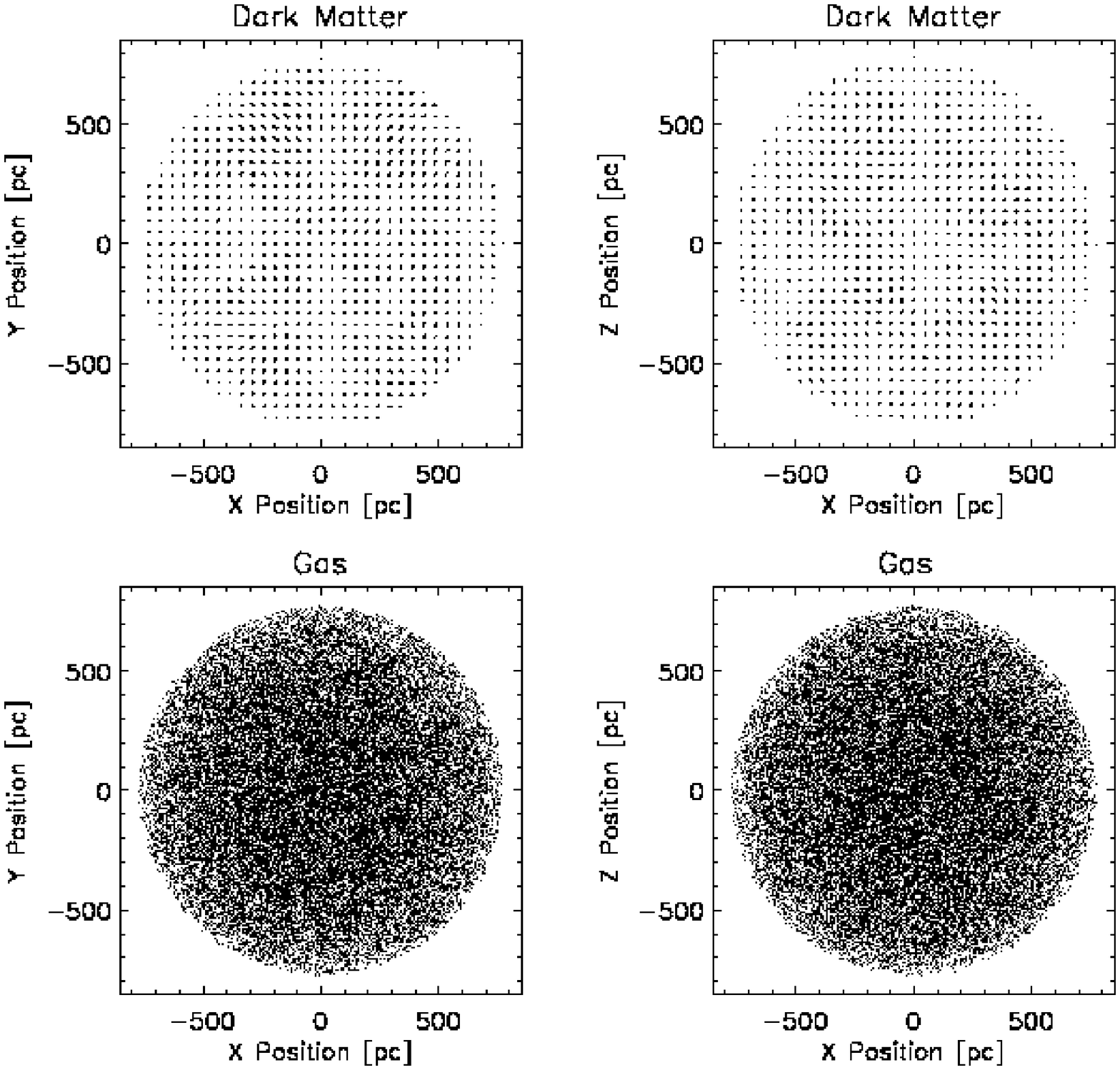,width=8.4cm,height=7.56cm}
\figcaption{
Run A: Initial configuration for top-hat collapse at $z_{i}=100$.
The halo has a total mass of
$10^{8}M_{\odot}$, and is endowed with a Hubble expansion such
that virialization occurs at $z_{\rm vir}\simeq 10$.
{\it Top row:} The DM particles are perturbed from a regular grid
according to $P(k)\propto k^{-2.5}$.
{\it Bottom row:} The gas particles are placed at random, and comprise a
mass fraction of 15\%.
{\it Left panels:} Face-on view.
{\it Right panels:} Edge-on view.
\label{fig1}}
\end{center}

\begin{center} 
\epsfig{file=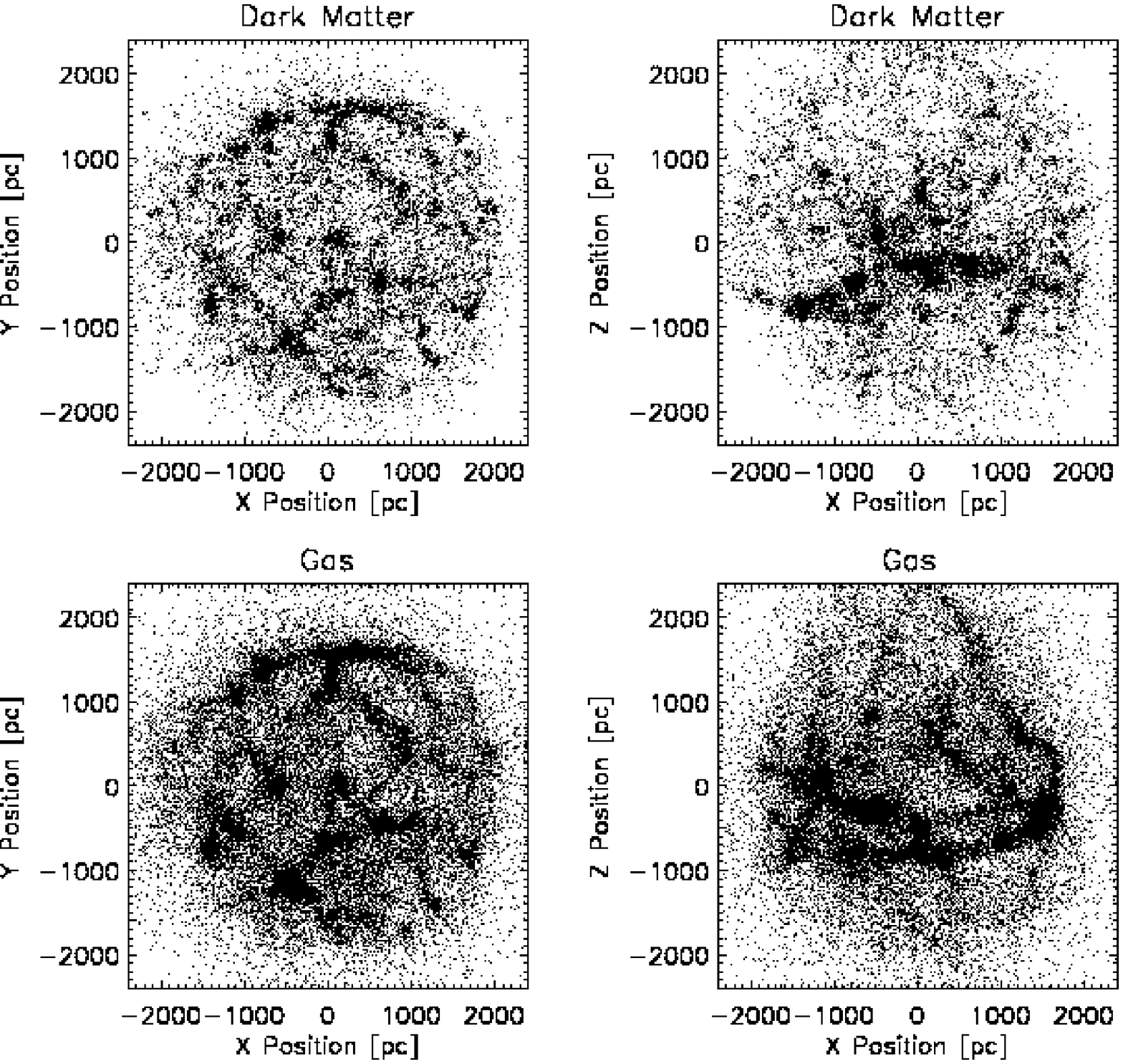,width=8.4cm,height=7.56cm}
\figcaption{
Run A: Morphology at $z\sim 13$.
The manner of presentation is the same as
in Figure 1, but with a box size of $\sim 5$~kpc.
The DM has developed significant substructure, and the baryons are just
beginning to fall into the corresponding potential wells.
\label{fig2}}
\end{center}

\begin{center} 
\epsfig{file=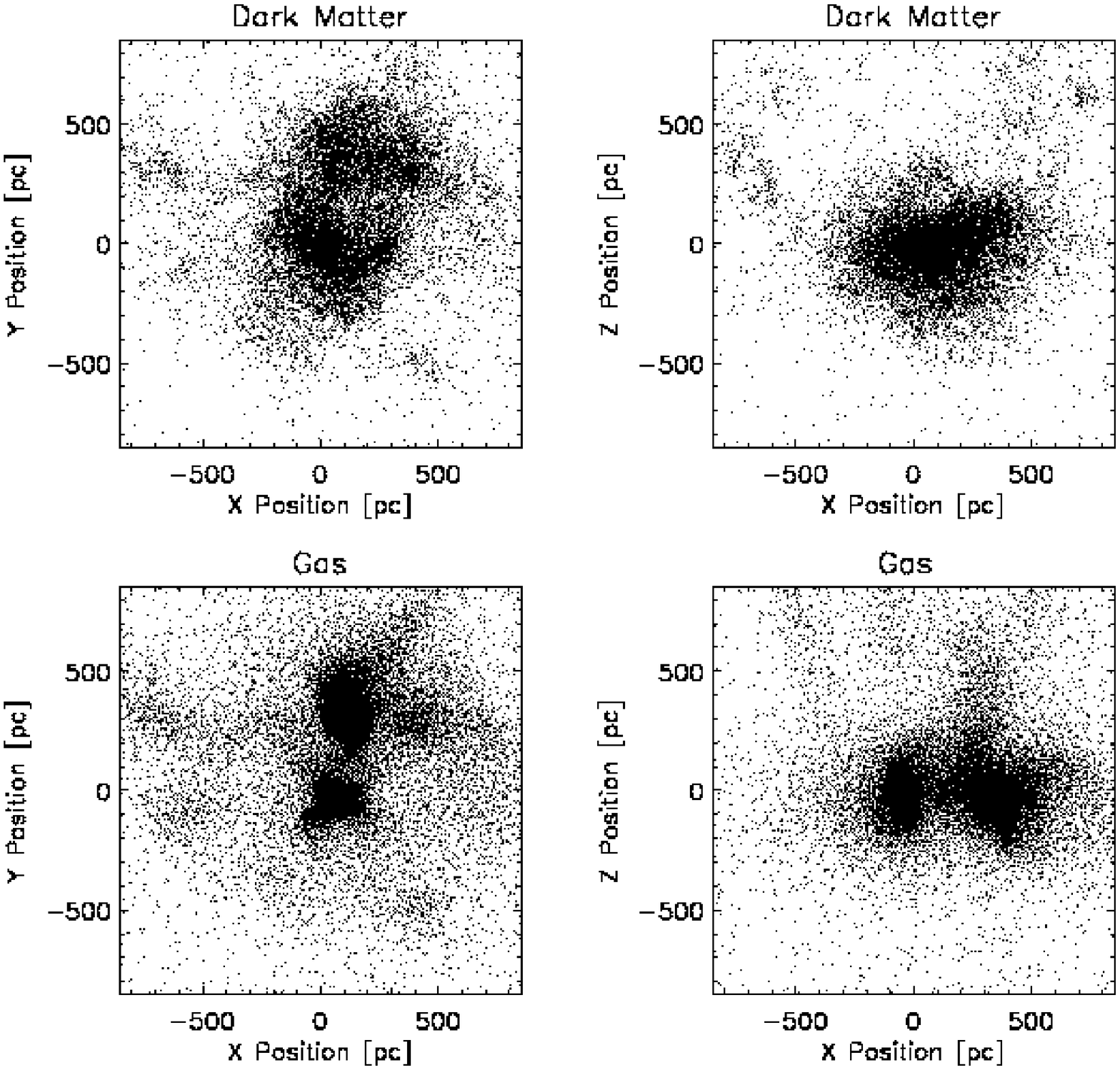,width=8.4cm,height=7.56cm}
\figcaption{
Run A: Morphology at $z=10.3$.
The convention in Fig. 1 is adopted for the rows and columns.
The box size is $\sim 2$ kpc.
The DM is in the process of undergoing violent relaxation with the concurrent
smoothing out of substructure. 
The gas has settled into the center
of the DM potential well. 
\label{fig3}}
\end{center}

Figure 2 shows the system at $z\sim 13$, briefly after
turnaround.  The DM component has developed a marked substructure in
response to the imprinted perturbations, and the baryons have begun to
fall into the deepest DM potential wells. Eventually, close to the
redshift of virialization, $z_{\rm vir}\simeq 10$, the DM is undergoing
violent relaxation (Lynden-Bell 1967), resulting in an approximate balance
between kinetic and gravitational potential energy.

\begin{center} 
\epsfig{file=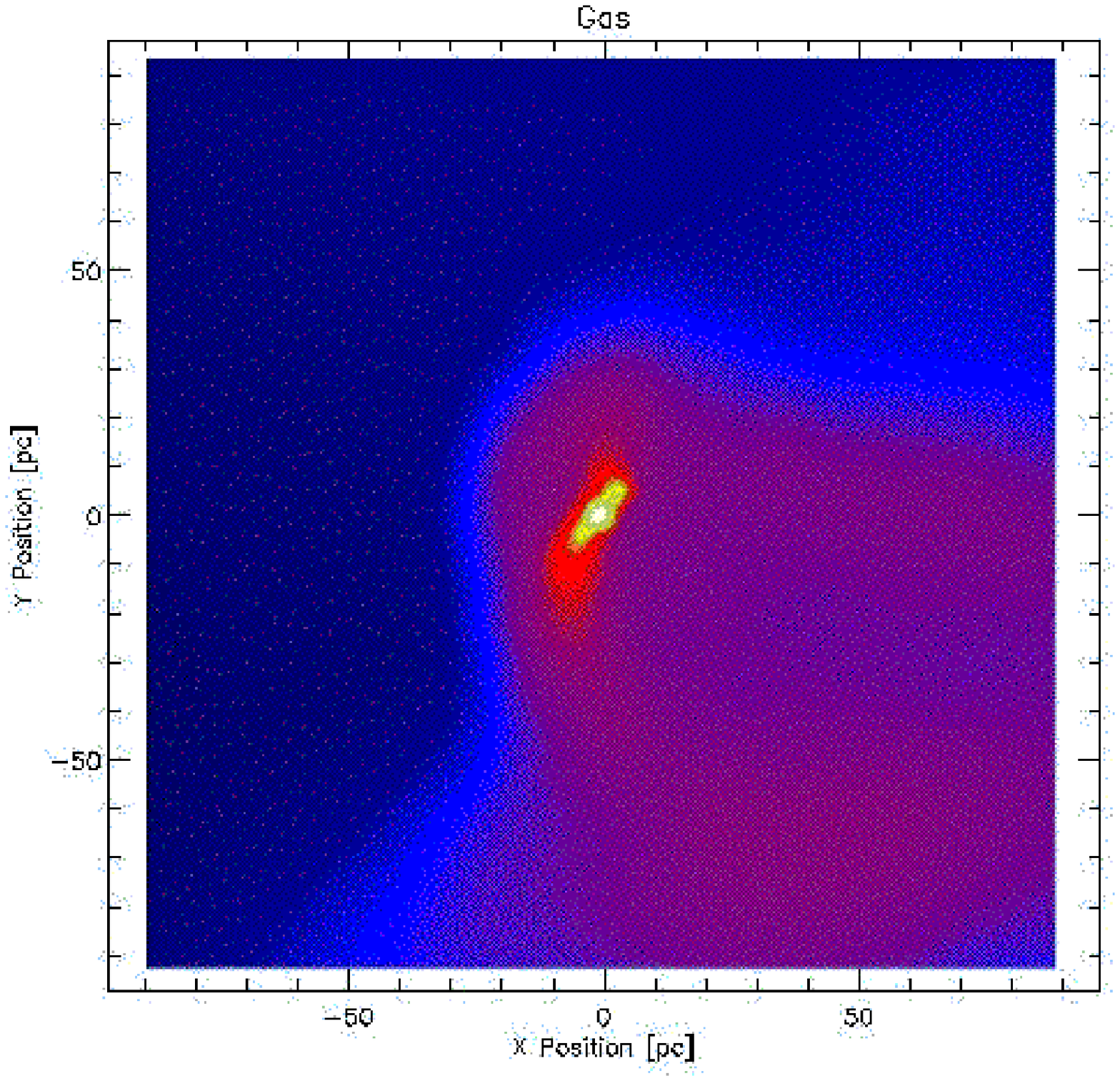,width=8.4cm,height=7.56cm}
\figcaption{
Run A: Central gas density at $z=10.3$, in the case with zero initial spin.
The box size is 200 pc. One compact object has formed in the center with
a mass of $2.7\times 10^{6}M_{\odot}$ and a radius of $\la$1 pc.
\label{fig4}}
\end{center}

The baryons, on
the other hand, dissipatively settle into the center of the DM halo
(see Figure 3). Focusing on the innermost $\sim 200$~pc of the
simulation box, Figure 4 displays the central distribution of the gas
density.  At this stage in the simulation, a high-density sink
particle of mass $\sim 3\times 10^{6}M_{\odot}$ and radius $\la 1$~pc
has been created.  Subsequently, the sink particle grows in mass by
continuous accretion from the surrounding diffuse gas.

\begin{center} 
\epsfig{file=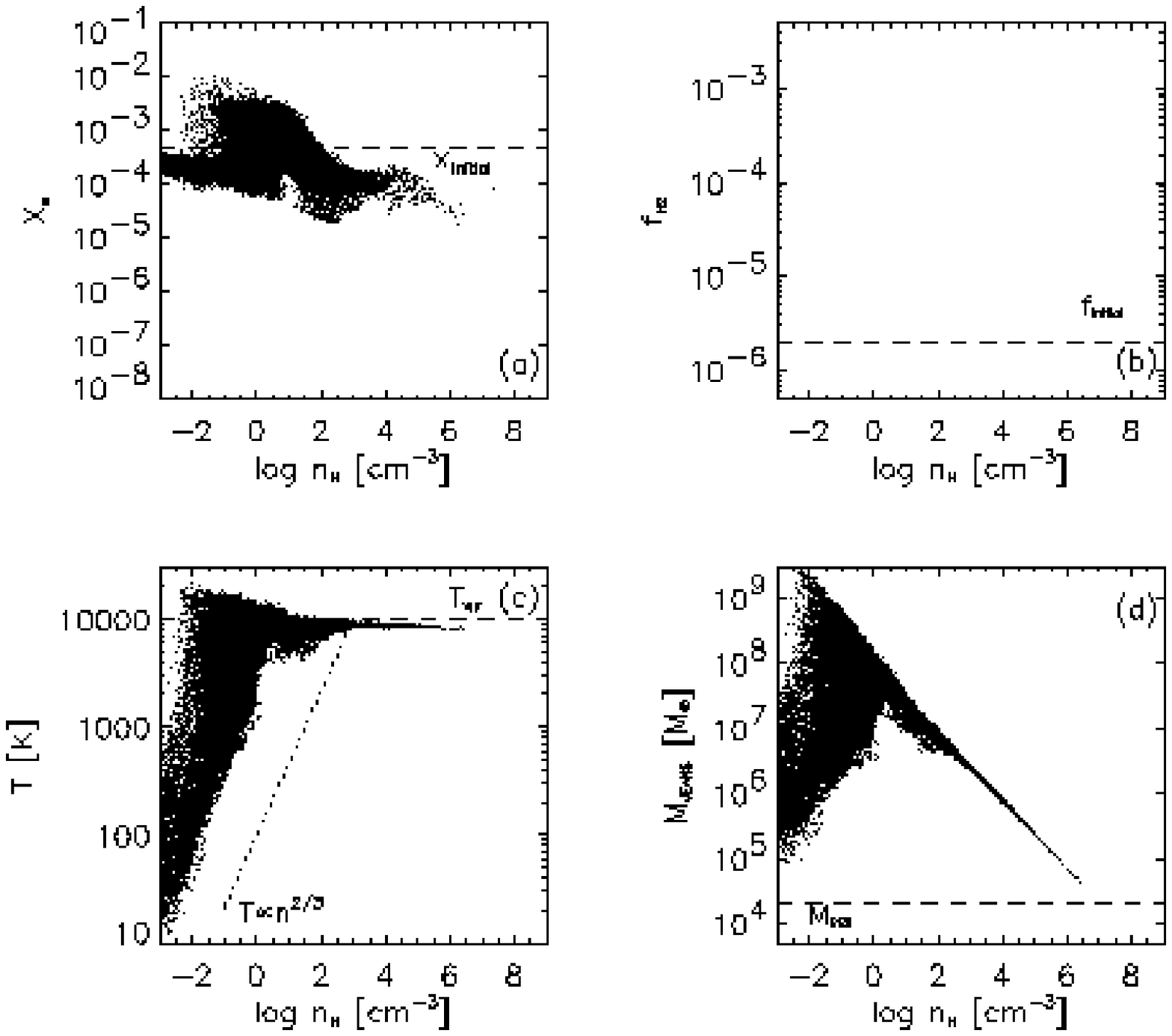,width=8.4cm,height=7.56cm}
\figcaption{
Run A: Gas properties at $z=10.3$.
{\bf (a)} Free electron abundance vs. hydrogen number density (in cm$^{-3}$).
{\bf (b)} Hydrogen molecule abundance vs. number density.  In this
run, we do not allow for the formation of H$_{2}$, but compare with
the corresponding panel in Fig.~7.  {\bf (c)} Gas temperature
vs. number density. At densities below $\sim 1$ cm$^ {-3}$, the gas
temperature rises because of adiabatic compression until it reaches
the virial value of $T_{\rm vir}\simeq 10,000$ K.  At higher densities,
cooling due to atomic H keeps the gas nearly isothermal.  {\bf (d)}
Jeans mass (in $M_{\odot}$) vs. number density. The Jeans mass reaches
the resolution limit of the simulation, $M_{\rm res}\simeq 20,000 M_{\odot}$,
for hydrogen number densities close to the merging threshold of $n_{\rm
th}=10^{7}$ cm$^{-3}$.
\label{fig5}}
\end{center}

To gain further insight into the physics of the simulation, we
consider the thermal and chemical properties of the gas, as
illustrated in Figure 5.  We show the free-electron abundance,
temperature, and Jeans mass as a function of gas density for every SPH
particle. This manner of presentation contains an additional dimension
of information, concerning the overall timescale of evolution: when
the evolution proceeds slowly, particles tend to pile-up in the
respective plots, whereas only a few particles populate regions in
parameter space where the evolution is fast.  As can be seen in panel
(d), a gas density of $n\sim 10^{2}{\rm cm}^{-3}$ and a Jeans mass of
$M_{J}\sim 10^{6}M_{\odot}$ approximately mark the transition between
a phase of slow and fast evolution. Physically, this transition
corresponds to the onset of runaway collapse when the clump mass is $M\ga
M_{J}$.

In Figure 6, we show the central gas configuration for the simulation
where the initial spin of the system is nonzero (Run B). In this case,
a binary system of clumps has formed with a separation of $\sim
10$~pc. These two clumps individually have properties similar to the
one clump formed in Run A. The binary is rapidly drawn into the center
of the halo by dynamical friction which operates on a timescale
$t_{\rm df}\sim 0.1 r^{2}v_{c}/(G M_{\rm Cl})\sim 10^{7}$~yr (Binney
\& Tremaine 1987). Here, $v_{c}$ is the circular velocity at radius
$r$, and $M_{\rm Cl}$ is the clump mass. This timescale is short
compared to the Hubble time at $z\sim 10$, $t_{H}\simeq
10^{8}$~yr. Such a system of two compact objects is expected to
efficiently radiate gravitational waves that could be detected with
the planned {\it Laser Interferometer Space Antenna\footnote{See
http://lisa.jpl.nasa.gov/}} (LISA) (Wyithe \& Loeb 2003b).

\begin{center} 
\epsfig{file=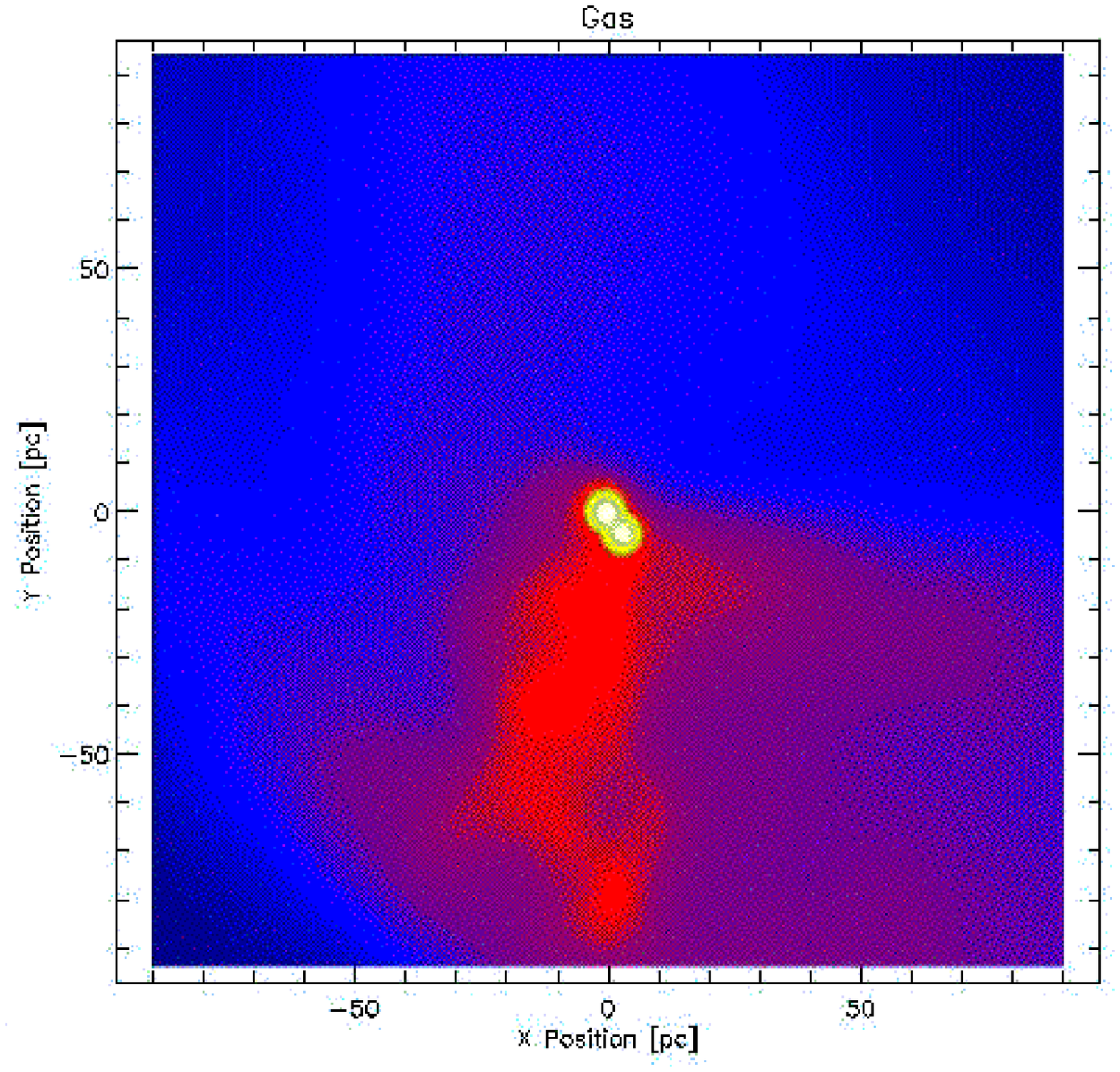,width=8.4cm,height=7.56cm} \figcaption{ Run B:
Central gas density at $z\simeq 10$, in the case with an initial spin
parameter of $\lambda=0.05$.  The box size is 200 pc. Here, two
compact objects have formed in the center with masses of $2.2\times
10^{6}M_{\odot}$ and $3.1\times 10^{6}M_{\odot}$, respectively, and
radii $\la$1 pc.
\label{fig6}}
\end{center}

Massive clumps, either in isolation or as a binary, are able to form
in both runs A and B, which correspond to different values of initial spin.
The result that such a compact object can form even in a halo with average
initial spin (run B), is seemingly at odds with the earlier investigation
by Eisenstein \& Loeb (1995), who have argued that only the lowest-spin
cosmological perturbations can harbor seed BHs. The gas that is eventually
incorporated into the two massive clumps in run B must have been able to
efficiently lose angular momentum. It is crucial that the evolution
in the $\lambda=0.05$ case leads to the formation of a binary system.
Tidal torques can then transfer much of the angular momentum from the gas
around each forming clump to the orbital motion of the system, similar
to the case of present-day star formation in a clustered environment
(e.g., Larson 2002; Bate et al. 2003). We plan to more fully address
the complex physics of angular momentum transport in virializing DM halos
in future work.

This efficient mechanism of
assembling a large amount of gas into a compact ($\la 1$~pc)
configuration crucially depends on the suppresion of star formation in
the collapsing halo. In Run C, which otherwise has the same initial
conditions as Run A, we now allow for the formation of hydrogen
molecules. Figure 7 (which should be compared to Fig.~5 of Run A),
shows that the thermal evolution of the gas in this case proceeds very
differently from Runs A and B. Here, the gas is able to cool to
temperatures $T\ga 200$~K due to the presence of H$_{2}$.  Provided
that H$_2$ formation is not suppressed, the properties of primordial,
metal-free gas are rather similar in systems of mass $\sim 10^{8}
M_{\odot}$, and in the smaller halos of mass $\sim 10^{6}M_{\odot}$
that are predicted to host the very first stars at $z\simeq 20 - 30$
(see Fig.~10 in Bromm et al. 2002).

The thermal evolution is clearly reflected in the morphology.  Figure
8 shows the projected gas density. Due to efficient cooling, the gas
can readily fragment and subsequently undergo runaway collapse to form
stars already shortly after turnaround.  This star formation activity
will tend to prevent the assembly of gas in compact clumps in two
ways. First, the Population~III stars are expected to be short-lived,
with lifetimes of $\sim 3\times 10^{6}$~yr, and they will exert a
strong negative feedback on their surroundings upon their death as
supernovae (SNe).  Second, the SNe will disperse the nucleosynthetic
products from the first generation of stars into the remaining gas of
the halo. 
Henceforth, the enriched gas will be able to cool and form
stars regardless of whether H$_{2}$ is present or not.
Since the photodissociation of H$_{2}$ is now no longer able to
limit the gas condensation (see \S~2), the star formation rate
is expected to increase substantially (e.g., Nishi \& Tashiro 2000; 
Bromm \& Clarke 2002).

\begin{center} 
\epsfig{file=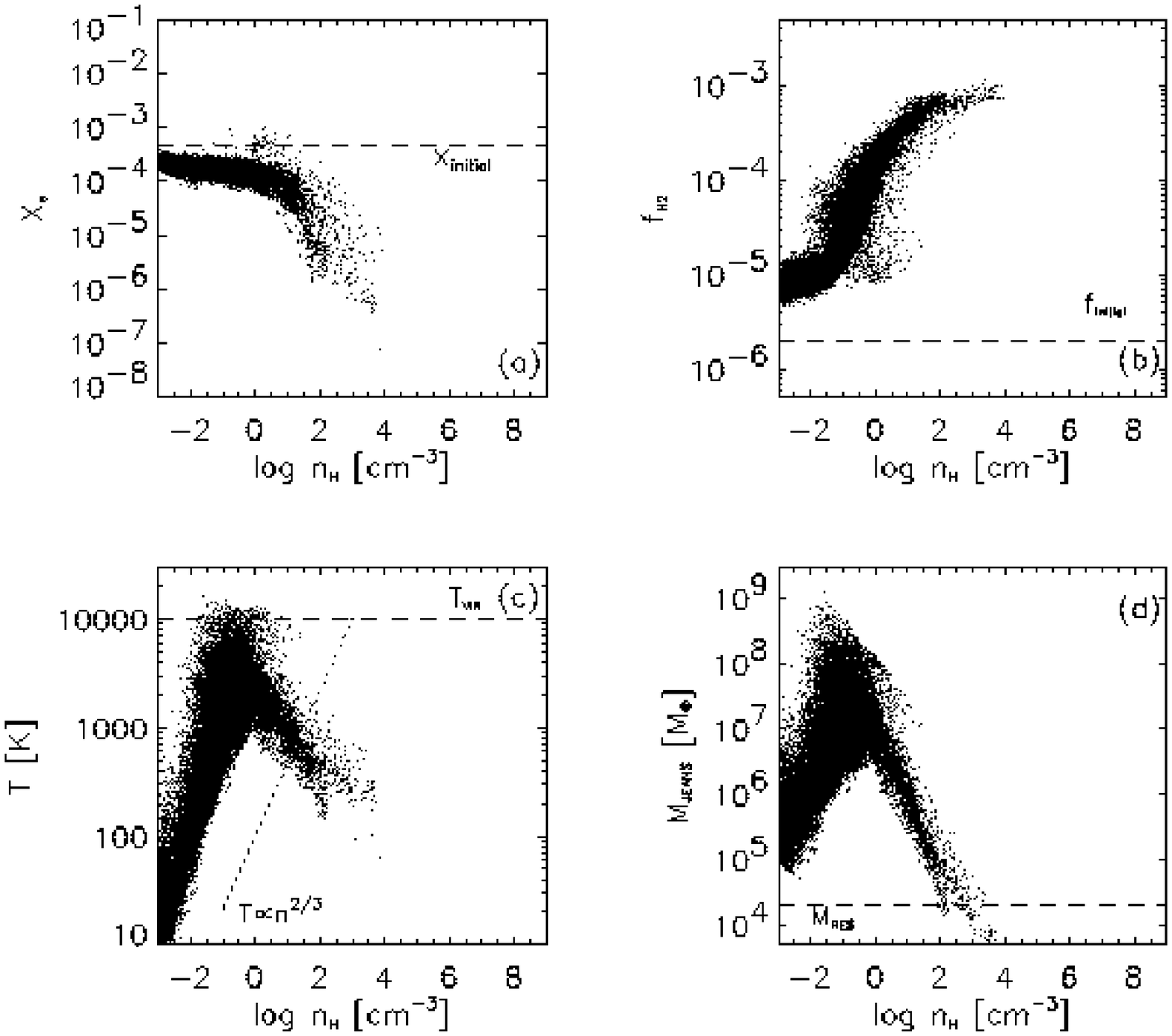,width=8.4cm,height=7.56cm} \figcaption{ Run C:
Gas properties at $z\sim 12$.  {\bf (a)} Free electron abundance
vs. hydrogen number density (in cm$^{-3}$).  At densities exceeding
$n_{\rm H}\sim 10^{3}$ cm$^{-3}$, recombination is very efficient, and
the gas becomes almost neutral.  {\bf (b)} Hydrogen molecule abundance
vs. number density.  In this run, we do allow for the formation of
H$_{2}$. The thermal behavior of the gas is markedly different from
Runs A and B, where molecule formation was suppressed.  {\bf (c)} Gas
temperature vs. number density. In contrast to Runs A and B, cooling
due to H$_{2}$ drives the temperature down at higher density, to
values of $\sim$ 200 K.  {\bf (d)} Jeans mass (in $M_{\odot}$)
vs. number density. The Jeans mass reaches the resolution limit of the
simulation, $M_{\rm res}\simeq 20,000 M_{\odot}$, for densities close to
the merging threshold of $n_{\rm th}=10^{4}$ cm$^{-3}$.
\label{fig7}}
\end{center}

The massive clumps formed in runs A and B are evidently not yet
BHs. To examine the subsequent dynamics of the clump, we have
resimulated the evolution of a typical clump with higher resolution in
order to reach smaller spatial scales. We report on this simulation next.

\subsection{Small-scale evolution}

As a representative example, we have selected Run~A for initializing
the refined simulation. We focus on the central region, as shown
in Figure~4, and resample the gas within a radius of $\sim 5$~pc
from the density maximum. The initial moment of the resampled run
is chosen to be somewhat earlier than in Figure~4, corresponding
to $z\simeq 10.6$, slightly before a sink particle was created in
the coarse-grain simulation. In the resampled region, the baryons
comprise a mass of $\sim 6\times 10^{5} M_{\odot}$, and are now
the dominant component, compared to only $\sim 10^{5}M_{\odot}$ in
dark matter.

After $\sim 10^{4}$~yr, we reach the end-state of our refined
simulation, at which point the Jeans mass again becomes comparable to
the resolution limit. The central gas cloud is in a state of
free-fall, and we do not see any signs of further
sub-fragmentation.
We have verified that fragmentation is not artificially suppressed
in the resampling process. To this extent, we have carried out
a fiducial simulation that does lead to fragmentation at some point
in the evolution, and a comparison calculation where we resample
the fluid {\it before} fragmentation occurs. We find that the resulting
fragmentation pattern is very similar in the two cases, as is 
physically expected.

The innermost region, of size $\la 0.1$~pc,
comprises $\sim 10^{4}M_{\odot}$ in gas with densities in
excess of $10^{9}$~cm$^{-3}$. If we were to continue the simulation
further in time, the amount of gas residing inside the central $\sim
0.1$~pc would rapidly increase. Indeed, the mass of the sink particle
formed in the large-scale simulation of Run~A, of order a few
$10^{6}M_{\odot}$, is indicative of the total baryonic mass that will
end up in the central compact object.  At this stage, this object is
characterized by a ratio $E_{\rm rot}/|E_{\rm grav}|\simeq 0.5$, where
$E_{\rm rot}$ and $E_{\rm grav}$ are the rotational and gravitational
energies of the central compact object, respectively.
The angular momentum present in the central clump must have arisen
through torques from the clumpy DM distribution during the relaxation
process, as run~A has zero initial spin. Again, we plan to revisit this
issue in our planned work on angular momentum transport in collapsing DM
halos.

\begin{center} 
\epsfig{file=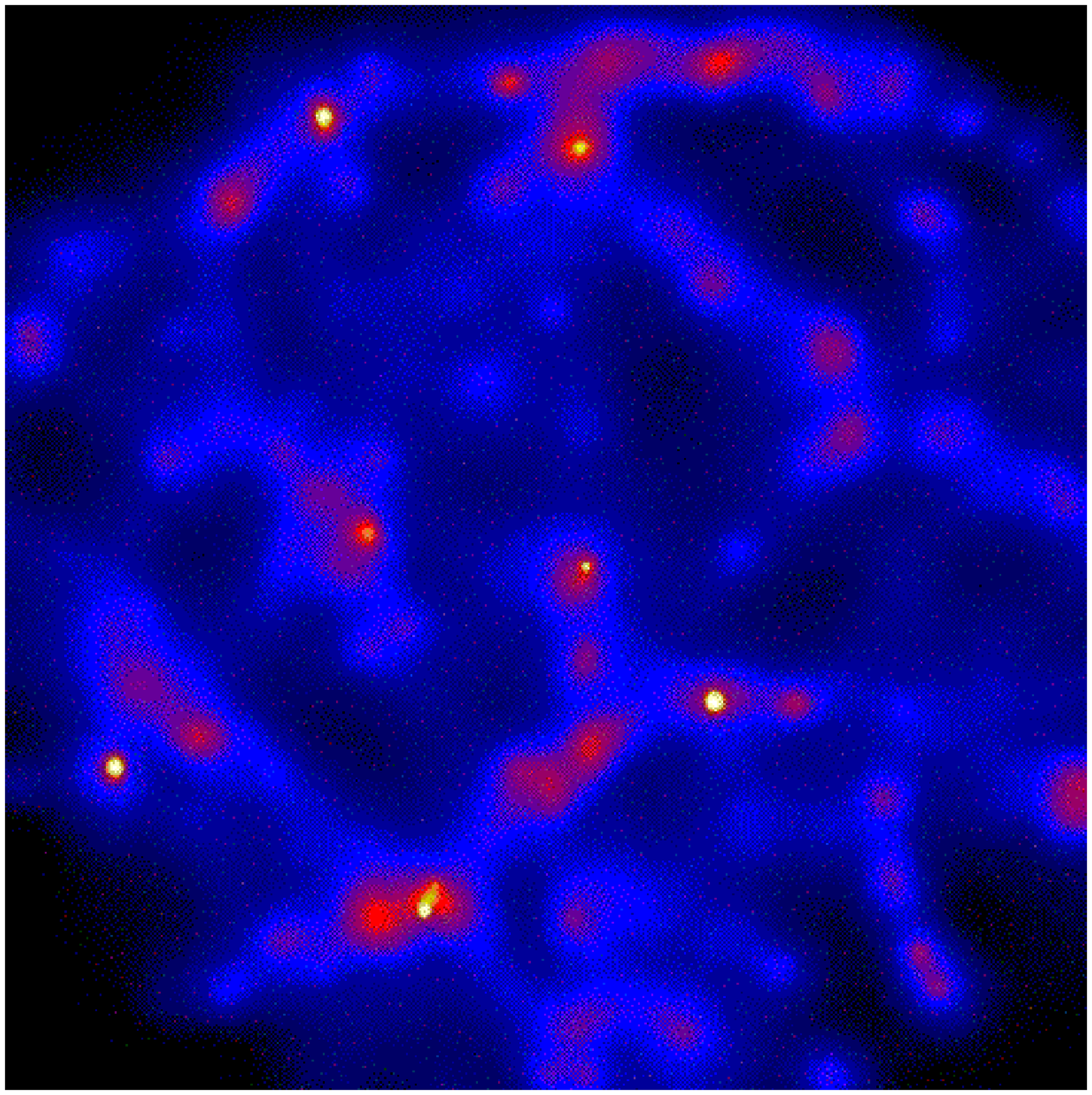,width=8.4cm,height=7.56cm}
\figcaption{
Run C: Morphology at $z\sim 12$.
Shown is the projected gas density in the x-y plane.
The box size is $\sim 3$ kpc.
Here, the gas readily fragments and subsequently undergoes runaway collapse
to form stars due to the efficient cooling provided by H$_{2}$.
\label{fig8}}
\end{center}

The crucial question now is: {\it What is the further fate of the
central object?} In particular, do we expect the cloud to fragment,
and eventually form a central stellar cluster? Or else, will
fragmentation be inhibited, as in the earlier, large-scale evolution
described in the previous section? In the latter case, the additional
question arises (Loeb \& Rasio 1994) whether the gas cloud will
continue to collapse directly to a black hole or else settle
temporarily into a pressure-supported configuration of a supermassive
star (see \S 17 in Shapiro \& Teukolsky 1983)? We can gain important
physical insight into these questions by comparing three important
timescales in the problem, namely: the free-fall time, $t_{\rm ff}$,
during which dynamical equilibrium is established; the cooling time,
$t_{\rm cool}$, during which the gas may radiate its thermal energy as
it contracts and heats to higher temperatures; and the viscous time,
$t_{\rm vis}$, during which angular momentum is transported.  The
cooling time remains relatively short and allows continued collapse
until the cloud becomes optically thick to the radiation it produces.
Spherically-symmetric calculations indicate that when the particle
density in the contracting gas clump rises above $\sim 10^{17}~{\rm
cm^{-3}}$, the ionization fraction increases sharply by 7 orders of
magnitude to a value $x_e\ga 0.1$ (see Fig. 4 in Omukai 2001). A clump
of mass $\sim 2\times 10^6M_\odot$ achieves this dense phase when its
radius $R$ shrinks to $(R/10^{16}~{\rm cm})\equiv R_{16}\la 0.5$,
about two orders of magnitude smaller than the value at the end of our
simulations. When $R_{16}\la 1$, the optical depth of the cloud to
Thomson scattering by free electrons is $\tau \sim 10^7R_{16}^{-2}$
and so the cooling time is given by the photon diffusion time $t_{\rm
cool}\sim \tau R/c\sim 10^5~{\rm yr}~ R_{16}^{-1}$.  The free fall
time, $t_{\rm ff} \sim (G\rho)^{-0.5} \sim 1~{\rm yr}~R_{16}^{1.5}$,
is much shorter than this cooling time and so the gas may settle
into a radiation-pressure supported configuration resembling a
rotating supermassive star (SMS).  The viscous time is shorter than
the cooling time, $t_{\rm vis}\la 4\times 10^4~{\rm yr}~
R_{16}^{0.5}(P_{\rm gas}/P_{\rm tot})\alpha^{-1}$, where $\alpha\la 1$
is a dimensionless measure of the viscosity coefficient (Shakura \&
Sunyaev 1973; \S 14.5 in Shapiro \& Teukolsky 1983) in terms of the
total (gas$+$radiation) pressure, $P_{\rm tot}=(P_{\rm gas}+P_{\rm
rad})\gg P_{\rm gas}$. The resulting viscous dissipation will likely
heat the optically-thick gas close to its virial temperature since the
rotational energy is substantial at the end of our simulation.
Combined with additional heating from adiabatic compression, the
collapse may be halted as soon as the gas becomes highly
optically-thick and is unable to cool efficiently.  We therefore
conclude that if the simulated gas clouds continue to shrink during
the early optically-thin (efficient cooling) regime by about two
orders of magnitude in radius (as they do earlier in the simulations),
then they will likely form a rotating SMS.

Recent fully-relativistic calculations of the evolution of rotating
SMSs have predicted the collapse into a massive BH (Baumgarte \&
Shapiro 1999; Shibata \& Shapiro 2002). Under a wide range of initial
conditions, a substantial fraction of the mass of the SMS ($\sim
90\%$), is expected to end up in the BH.

\section{EFFECT OF A PHOTODISSOCIATING BACKGROUND}

We now return to the question: {\it What is the required level of the
LW background to prevent H$_{2}$ from forming when self-shielding is
taken into account?}  To answer this question, we have carried out a
series of simulations where we allow H$_{2}$ to form self-consistently
in the presence of a soft UV background. We have determined $J_{\rm
crit}$, the critical flux above which H$_{2}$ is not able to form
throughout the simulation volume, for the three spectra described in
\S~3.3. For both the power-law spectrum, as well as the thermal
spectrum with $T_{\ast}=10^{5}$~K, we find: $J_{\rm crit}\ga
10^{5}$. This value is larger than the expected background close to
reionization (see \S~2), and it would be difficult to prevent H$_{2}$
cooling for these two spectra.

A much lower background level, however, suffices in the case of a
thermal spectrum with $T_{\ast}=10^{4}$~K, for which we find: $J_{\rm
crit}\la 10^{3}$, close to the predicted UV background at $z\sim
10$. In Figure~9 we show the timescales that are relevant to
understand the abundance of H$_{2}$ in the simulation with
$J_{21}=10^{3}$. It is evident that even when self-shielding is
included, the H$_2$ formation time, $t_{\rm form}$, is longer than the
H$_2$ destruction time, $t_{\rm dest}$ throughout the simulation.
Once the gas has reached a density of $n_{\rm H}\ga 10^{3}~{\rm
cm^{-3}}$, collisional dissociation becomes very effective in
destroying H$_{2}$ at a gas temperature of $10^4$K. The
ever-increasing column densities, and the resulting level of
self-shielding, have therefore no effect on the molecule abundance at
densities high enough for collisions to become important.

Using an idealized one-zone model to describe the dynamics of a
collapsing primordial cloud, but otherwise implementing the relevant
radiative processes in a sophisticated way, Omukai (2001) has
determined the critical UV flux in the case of a Planck spectrum with
$T_{\ast}=10^{4}$~K to be: $J_{21} \sim 10^{3}$, in good agreement
with our numerical simulations.

\begin{center} 
\epsfig{file=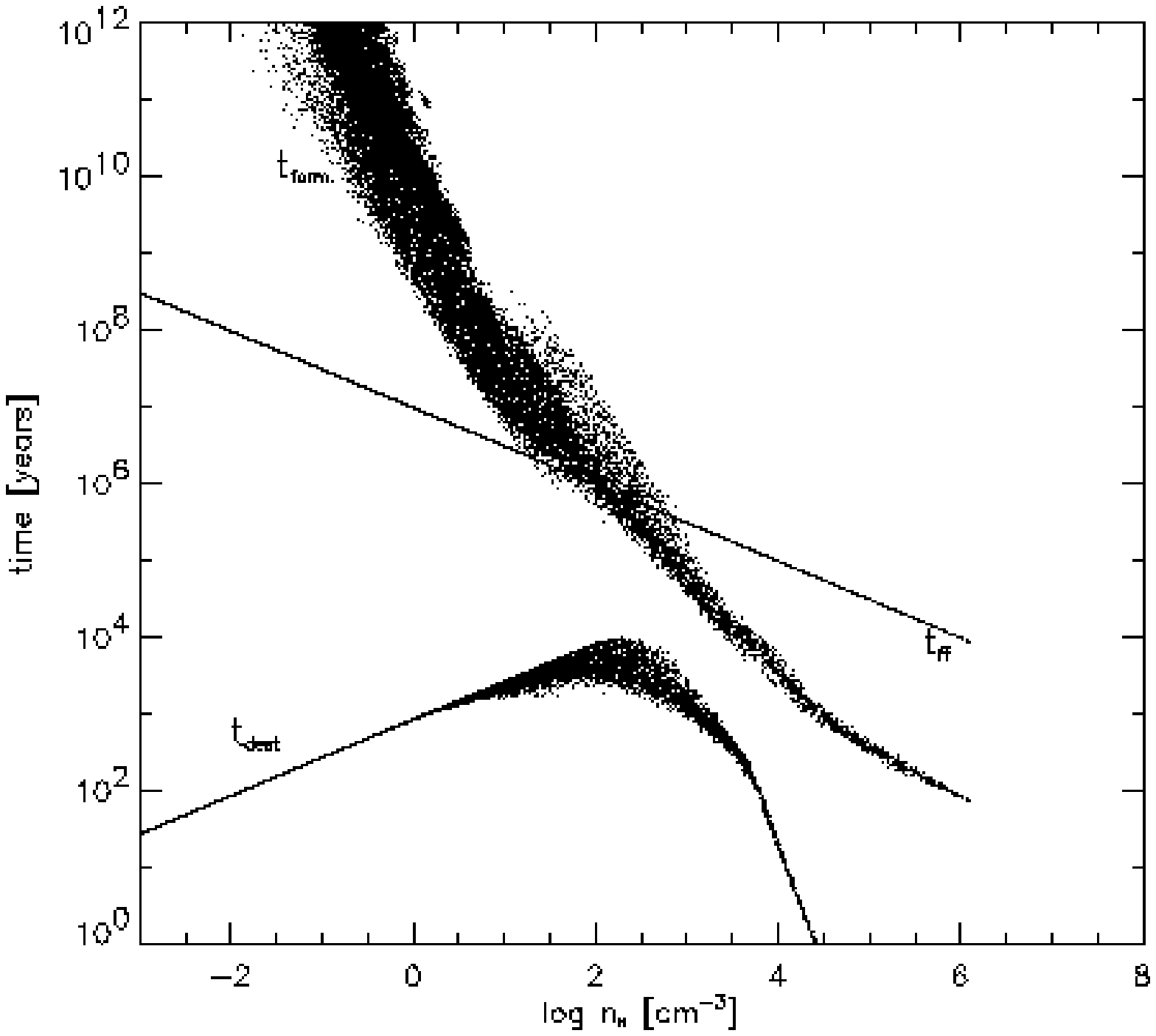,width=8.4cm,height=7.56cm} 
\figcaption{ Run D: Important timescales at $z=10.6$ for a background
stellar spectrum with $T_{\ast}=10^4$~K.  {\it Solid line:} Free-fall
time vs. hydrogen number density (in cm$^{-3}$).  {\it Dots:} H$_{2}$
formation and destruction times vs. $n_{\rm H}$ for every SPH particle in
the simulation.  The LW flux is normalized to $J_{21}=10^{3}$ at the
Lyman limit.  The destruction time increases with density up to
$n_{\rm H}\sim 10^{2}$ cm$^{-3}$ due to the effect of self-shielding.
H$_{2}$ is never formed efficiently, however, as $t_{\rm form}>t_{\rm
dest}$ throughout the simulated volume. At $n_{\rm H}\ga 10^{3}$ cm$^
{-3}$, H$_{2}$ is readily destroyed via collisions, and $t_{\rm dest}$
decreases again with density.
\label{fig9}}
\end{center}

Both these estimates crucially depend on the presence of a {\it
stellar-like} radiation background. The formation of the first
quasars, therefore, might well require an earlier epoch of star
formation. The old puzzle of whether quasars or stars were the first
luminous objects in the universe (e.g., Silk \& Rees 1998) would then
be answered in favor of the latter.

\section{SUMMARY AND CONCLUSIONS}

Our numerical simulations show that metal-free dwarf galaxies at
redshifts $z\sim 10$ whose cooling is dominated by atomic transitions
(with a virial temperature just above $10^4$K) tend to form a central
condensation consisting of one or two clumps which contain $\ga 10\%$
of the total baryonic mass of the galaxy.  As long as H$_{2}$
formation is suppressed, these massive clumps do not fragment but
rather cool and continue to collapse isothermally at a temperature of
$\sim 10^4$K.  We have simulated the collapse to the
point where $\ga 10^6M_\odot$ of gas condense to a scale $\la
1$pc. At the end of our simulation the clump maintains a nearly
free-fall collapse and shows no signs of fragmentation. We expect the
collapse to continue until the gas cloud becomes optically thick to
Thomson scattering and its cooling time is much longer than the
infall time. At this stage the collapse may be halted due to viscous
dissipation of rotational energy, and a rotating SMS is likely to
form. After shedding a small fraction of its mass, the SMS will
collapse to a SMBH (Baumgarte \& Shapiro 1999; Shibata \& Shapiro
2002).

The possible existence of a supermassive star as an intermediate stage
in the formation of the first quasars can be tested by direct
observations.  The SMS spectrum is expected to be close to thermal at
an effective temperature of $\sim 10^5$~K (Bromm et al.  2001b). The
SMS mass can be inferred from its bolometric luminosity, given that
the latter equals the Eddington value, $L_{\rm Edd}=1.4\times
10^{44}~{\rm erg~s^{-1}}(M/10^6M_\odot)$.  The emission redshift can
be easily derived from the Gunn-Peterson trough produced by Ly$\alpha$
absorption of the neutral IGM (Gunn \& Peterson 1965).  Our model
predicts that the emission spectrum of the SMS would show only
hydrogen and helium lines but no metal lines (such as the broad
emission lines observed for quasars at lower redshifts), since metal
enrichment would signal small scale fragmentation into stars and would
in turn enable the production of molecules on dust grains.  However,
all the above characteristics are common to clusters of Population~III
stars (Bromm et al. 2001b). To distinguish between a SMS and the
extended image of a star cluster requires high angular resolution,
possibly achievable with the {\it James Webb Space
Telescope}\footnote{See http://ngst.gsfc.nasa.gov/}. Alternatively, it
is possible to make use of the high probability of gravitational
lensing by intervening galaxies out to $z\sim 10$ (Barkana \& Loeb
2000; Wyithe \& Loeb 2002a,b; Comerford, Haiman, \& Schaye 2002).  In
particular, gravitational microlensing by stars within the lens
galaxies (Wyithe \& Loeb 2002b) is only possible for a SMS and not for
a star cluster whose extent is much larger than the Einstein radius of
a solar-mass star at a cosmological distance, $\sim 10^{-2}~{\rm pc}$.

Our calculations show that fragmentation into stars is suppressed
inside the above dwarf galaxies only if H$_2$ cooling is
negligible. The photodissociation of H$_2$ molecules requires a
background flux of UV photons below 13.6eV (to which the neutral IGM
is nearly transparent) that could possibly be produced by stars at $z\sim
10$, prior to reionization (see \S 2).

{\it What is the comoving density of the first generation of massive
BHs?}  At $z\sim 10$, the comoving density of halos with a virial
temperature $\ga 10^4$K is $\sim 5~{\rm Mpc}^{-3}$ (Sheth \& Tormen
1999).  Requiring that the total mass density of massive BHs would not
exceed the value observed in the local universe (Yu \& Tremaine 2002),
$\sim 2.5\times 10^5M_\odot~{\rm Mpc^{-3}}$, we infer that the average
BH mass per halo must be $\la 5\times 10^{4}M_\odot$, much smaller
than the characteristic mass of the gas clumps in our simulations.
This follows naturally from our expectation that only a small fraction
of all halos are both metal poor and exposed to a high level of UV
flux during their entire history (see \S 2 and \S 5).  Even though a
substantial fraction of the IGM volume is not enriched with metals at
$z\sim 10$ (Thacker et al. 2002; Furlanetto \& Loeb 2003), most
$2\sigma$ halos may have been enriched by stars that were formed
inside their hierarchical building blocks at an earlier cosmic time,
when the UV background was much lower. The metals may then facilitate
H$_2$ formation on dust particles as well as efficient cooling through
metal emission lines (both atomic and molecular), resulting in
further fragmentation.  In such a case, only a small minority of all halos
forms by direct accretion of pristine gas from the IGM, as we have
assumed in our simulations.

Alternatively, the final BH may contain a small fraction of the initial
mass of its parent clump of gas.  Since we did not follow the collapse
of each clump down to its Schwarzschild radius, we cannot be certain
of the final BH formation efficiency.  It is possible that
hydrodynamic or radiative feedback from a growing BH seed at the
center limits the final mass that it obtains, so that eventually most
of the mass of its parent clump is expelled in a wind due to energy
release near the center (e.g., Haehnelt, Natarajan, \& Rees 1998; Silk
\& Rees 1998).  The correlation between BH mass and the velocity
dispersion of its host galaxy as observed in the local universe
(Merritt \& Ferrarese 2001; Tremaine et al. 2002) would imply an
exceedingly low efficiency ($\la 10^{-4}$) for the BH formation
process in the dwarf galaxies under consideration (Wyithe \& Loeb
2002c). However, this phenomenological correlation is measured in
metal-rich galaxies where molecular cooling and star formation are
abundant, and it may not hold under the unusual physical conditions
inside the first generation of metal-free galaxies where the seeds for
the present-day population of BHs were produced.

An early formation of SMBHs would lead to a considerable rate of
detectable bursts of gravitational radiation from the coalescence of
BH binaries at high redshifts (Wyithe \& Loeb 2003b). In addition to
the commonly discussed origin of binaries in galaxy mergers (a
possible example of which has recently been observed in the nearby
galaxy NGC~6240 by Komossa et al. 2003), we have demonstrated that BHs
may form in binaries to start with, just as stars do.

\acknowledgments

We are indebted to Lars Hernquist for making available to us a version
of TREESPH. We also thank Rennan Barkana and the anonymous referee
for valuable comments on the manuscript. 
This work has been supported in part by NSF grants
AST-0071019, AST-0204514.  VB thanks the Institute for Advanced Study
for its hospitality during the completion of this work.  AL
acknowledges support from the Institute for Advanced Study at
Princeton and the John Simon Guggenheim Memorial Fellowship.


\clearpage

\end{document}